# Shear-induced polydomain structures of nematic lyotropic chromonic liquid crystal disodium cromoglycate

Hend Baza[a,b], Taras Turiv[b,c], Bing-Xiang Li[b,c], Ruipeng Li[d], Benjamin M. Yavitt[d,e], Masafumi Fukuto[d], and Oleg D. Lavrentovich*[a,b,c]

Lyotropic chromonic liquid crystals (LCLCs) represent aqueous dispersions of organic disk-like molecules that form cylindrical aggregates. Despite the growing interest in these materials, their flow behavior is poorly understood. Here, we explore the effect of shear on dynamic structures of the nematic LCLC, formed by 14wt% water dispersion of disodium cromoglycate (DSCG). We employ in-situ polarizing optical microscopy (POM) and small-angle and wide-angle X-ray scattering (SAXS/WAXS) to obtain independent and complementary information on the director structures over a wide range of shear rates. The DSCG nematic shows a shear-thinning behavior with two shear-thinning regions (Region I at $\dot{\gamma} < 1\ s^{-1}$ and Region III at $\dot{\gamma} > 10\ s^{-1}$) separated by a pseudo-Newtonian Region II ($1\ s^{-1} < \dot{\gamma} < 10\ s^{-1}$). The material is of a tumbling type. In Region I, $\dot{\gamma} < 1\ s^{-1}$, the director realigns along the vorticity axis. An increase of $\dot{\gamma}$ above 1 s$^{-1}$ triggers nucleation of disclination loops. The disclinations introduce patches of the director that deviates from the vorticity direction and form a polydomain texture. Extension of the domains along the flow and along the vorticity direction decreases with the increase of the shear rate to $10\ s^{-1}$. Above $10\ s^{-1}$, the domains begin to elongate along the flow. At $\dot{\gamma} > 100\ s^{-1}$, the texture evolves into periodic stripes in which the director is predominantly along the flow with left and right tilts. The period of stripes decreases with an increase of $\dot{\gamma}$. The shear-induced transformations are explained by the balance of the elastic and viscous energies. In particular, nucleation of disclinations is associated with an increase of the elastic energy at the walls separating nonsingular domains with different director tilts. The uncovered shear-induced structural effects would be of importance in the further development of LCLC applications.

.

## I. Introduction

Dynamics of nematic liquid crystals driven out of equilibrium by externally applied shear is a fascinating domain of soft matter physics. Over the last few decades, significant progress has been achieved in understanding shear-induced structures and rheology response of thermotropic low molecular weight nematics (LMWNs), nematics formed by surfactant-based wormlike micelles, viral suspensions, liquid crystal polymers (LCPs) and their solutions [1-3]. In LMWNs, the prime effect of shear flow on the system is through a spatially varying director $\hat{n}$ ($\hat{n} \equiv -\hat{n}$, $\hat{n}^2 = 1$) that specifies the average local molecular orientation. A combined effect of viscous, elastic, and surface torques produces a complex director field $\hat{n}(r)$ that often carries topological defects such as disclination loops, especially in the so-called tumbling nematics, in which the viscous torque is nonzero for any orientation of $\hat{n}$ [4].

At typical shear rates $\dot{\gamma}$, the degree of the orientational order, defined by the scalar order parameter $S$ of LMWNs, remains intact, since the molecular relaxation times are shorter than the time $1/\dot{\gamma}$ [3,5]. In LCPs, however, the relaxation times are long and flow can alter both $\hat{n}$ and $S$ [1,2]. LCPs are practically always tumbling; the occurrence of shear-induced singular topological defects-disclinations is a common scenario [6-11]. Although both $\hat{n}$ and $S$ can be altered by shear in LCPs, the flows are never sufficiently strong to change the length of the polymer chains since these are formed by strong covalent bonds.

There is a large class of self-assembled nematics, in which the building units are formed by weak non-covalent interactions. Among these are wormlike micellar nematics [12,13] and lyotropic chromonic liquid crystals (LCLCs) [14-16]. In wormlike micellar nematics, the building units are cylindrical aggregates formed by closely packed surfactant molecules that expose their polar heads towards the water (or other polar solvents). Extensive studies of rheological properties of wormlike micelles brought about a plethora of interesting effects, such as the shear-induced birefringent state in the isotropic solution accompanied by an increase of viscosity [17], formation of shear banding [18] in a biphasic region of the phase diagram [19], occurrence of damped director oscillations at intermediate shear rates in the nematic phase, attributed to tumbling [20], complete alignment of the micelles into a monodomain at high shear rates [21], or a shear-induced isotropic-to-nematic phase transition demonstrated both theoretically [22-25] and experimentally [26].

The nematic LCLC is formed by plank-like molecules that stack face-to-face on top of each, being attracted by noncovalent hydrophobic interactions, Fig. 1. In the resulting columnar aggregates, the molecular cores are stacked along the axis of the aggregate with a well-defined periodicity of about 0.34 nm. The columnar aggregates of LCLCs are polydisperse in length, with an average length that varies strongly with temperature and concentration [27]. The energy of scission, i.e. the energy needed to divide a chromonic aggregate into two is low, on the order of 10 $k_B T$ [15, 28-31]. The viscoelastic properties of LCLCs are much less understood as compared to those of other nematics.

The interest in the structural and rheological response of LCLCs to the applied shear is both practical and fundamental. LCLC materials can be used in applications such as sensors of microbial presence [32], templates for alignment of graphene layers [33], paintable polarizers [34, 35], organic field-effect transistors, [36] and optical compensators [16]. Preparation of the paintable polarizers, transistors, and optical compensators involves shear displacements of LCLCs with a simultaneous or subsequent drying. Although under strong shear deformations the LCLC aggregates align on average along the shear direction [37-44], one of the problems is the formation of a periodically modulated director field that can be related to both flow- and elasticity-triggered effects [45]. Cha et al [42] noticed that the combined shear and crystallization of LCLCs through water evaporation from isotropic solutions might result in the alignment of aggregates either parallel to the flow direction at high shear rates (above 100 s$^{-1}$) or perpendicular to it, at low shear rates on the order of 10 s$^{-1}$. Realignment of the chromonic aggregates from the flow-induced orientation parallel to the flow to a perpendicular orientation was reported also for the columnar phase of LCLCs by Crowley et al [46].

Fundamental understanding of how the LCLC structure responds to shear is practically absent. In particular, it is not known for certain which of these materials are flow-aligning and which are tumbling, whether the shear flow could trigger nucleation of disclinations, what are the differences in the viscous response of different phases of LCLCs, such as isotropic, nematic, and columnar phases. Because of the relative weakness of the aggregation forces in LCLCs, one might expect that shear can influence not only the director but also the scalar order parameter, since the aggregates could dissociate into smaller units or connect into longer ones, thus producing new effects not readily observed in covalently-bound systems but expected for wormlike micelles [47, 48]. The interest to explore the hydrodynamics of LCLCs is also caused by the fact that LCLCs are nontoxic to living organisms and can be interfaced with swimming organisms to produce a "living liquid crystal" [49], an experimental model of an active matter, in which activity, orientational order, and viscoelasticity are intimately interconnected. This interconnection allows one to use LCLCs as a structured environment to command dynamics of micro-swimmers such as motile bacteria [50-53].

The existing rheological studies of LCLCs focus mostly on the measurements of the average viscosity as a function of $\dot{\gamma}$ in cone-plate rheometers, reported for two typical nematic materials, namely, disodium cromoglycate (DSCG), Fig. 1a, [54] and Sunset-Yellow (SSY) [55]. Alternative approaches determine the effective viscosity through random Brownian [56-58] or externally directed motion of colloids [58-60] in LCLCs. Relaxation of magnetic field-induced twist deformations is used to determine the rotational (twist) viscosity of SSY [61]. Dynamic light scattering is employed to measure both the elastic and viscous coefficients corresponding to splay, bend, and twist deformations in DSCG [31]. In the case of SSY, polarizing Raman spectroscopy of the orientational order [62] lead to a conclusion that SSY is a tumbling nematic [63]; in which so-called tumbling parameter $\frac{\alpha_2 + \alpha_3}{\alpha_2 - \alpha_3}$ is smaller than 1; here $\alpha_2$ and $\alpha_3$ are the Leslie viscosities. However, in situ rheological x-ray scattering data are interpreted as the evidence that SSY is flow-aligning [44].

The objective of this study is to explore the director field response of the nematic LCLC DSCG under shear flow, by using plate-plate rheometry with an in-situ polarizing optical microscopy (POM) and small-angle and wide-angle x-ray scattering (SAXS/WAXS). The in-situ structural data are supplemented by the measurements of shear viscosity in a cone-plate geometry, which demonstrates shear-thinning behavior. This paper is organized as follows. First, we describe the materials and approaches used. Second, we present the measurements of average shear viscosity which point to three distinct regimes of response, two shear-thinning and one of apparent Newtonian behavior. Third, we perform a detailed analysis of POM textures and SAXS/WAXS data, supplemented by the measurements of the effective optical retardance to establish the cascade of structural transformations in the system as a function of the shear rate. At the lowest shear rates, in the shear-thinning Region I, $\dot{\gamma} < 1$ s$^{-1}$, the director realigns towards the vorticity axis, forming a log-rolling state. In the pseudo-Newtonian Region II, $1$ s$^{-1} \leq \dot{\gamma} \leq 10$ s$^{-1}$, the flow creates two distinct types of regions, one with the director still along the vorticity axis and another with the director tilted towards the shear plane, defined by the velocity **v** (the $x$-axis in Fig. 2a) and its gradient (the $z$-axis). As the shear rate increases, the elastic energy of director gradients associated with these domains becomes too high, and the stresses are released by nucleating singular disclinations of semi-integer strength. In Region III, the disclinations are progressively replaced with stripes in which the director, being predominantly along the flow, tilts left and right towards the vorticity axis. The observations prove the tumbling character of the DSCG nematic and are explained by considering the balance of viscoelastic forces.

## II. Materials and Methods

We explore the nematic phase of LCLC formed by 14wt% aqueous dispersion of DSCG, purchased from Spectrum Chemicals, Fig. 1. DSCG of 98% purity is dissolved in deionized (DI) water of resistivity $18.2 \times 10^4$ Ωm. The samples are prepared 48 hours before the experiments.

The optical birefringence $\Delta n = n_e - n_o$ of DSCG is negative [64], where $n_e$ and $n_o$ are the extraordinary and ordinary refractive indices. The absolute value of $\Delta n = -0.016$ at the wavelength 546 nm is measured in a planar cell, formed by two rubbed clean glass substrates separated by 5 μm spacers, using an LC PolScope and Berek compensator. This and all other experiments are performed at the temperature $T = 296$ K.

The effective shear viscosity as a function of the shear rate was measured in the cone-plate HAAKE MARS II rheometer (Thermo Scientific). The sheared volume is of a diameter 35 mm and a 4° cone angle. A Linkam Optical Shearing System CSS450 in a parallel disks geometry, Fig.2, is used to characterize the structural changes caused by the shear flows in both POM and SAXS/WAXS studies. The disks are parallel within 2 μm. The bottom plate rotates with a controllable angular velocity in the range (0.001-10) s$^{-1}$. In Fig. 2, we flip the geometry to simplify subsequent discussions. Both plates contain quartz windows for observations, of 2.5 mm in diameter, in transmitted mode, centered at 6.25 mm from the axis of rotation. Each plate is in thermal contact with an independently controlled pure silver heater utilizing platinum resistors with a sensitivity of 0.1°C. The gap $h$ between the two plates can be set in the range (5-2500) μm in 1 μm increment. The POM experiments are performed at $h=10\,\mu m$ while SAXS/WAXS data are collected for $h=(200-300)\,\mu m$. The plates are cleaned with DI water and acetone before each use. No alignment layers are used; the substrates of CSS450 yield degenerate tangential alignment of the director $\hat{\mathbf{n}}$. The optical characterization is performed for the shear rates in the range $0.75\,s^{-1} \leq \dot{\gamma} \leq 7500\,s^{-1}$, while SAXS/WAXS is performed for $0.04\,s^{-1} \leq \dot{\gamma} \leq 375\,s^{-1}$.

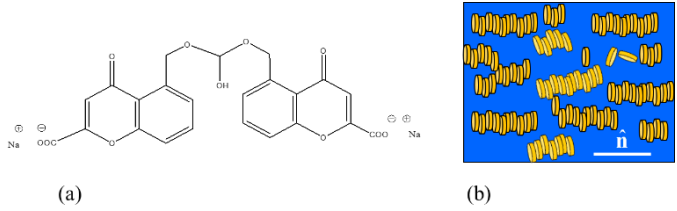

Fig. 1. (a) Molecular structure of DSCG; (b) aligned DSCG aggregates self-assemble into a nematic phase.

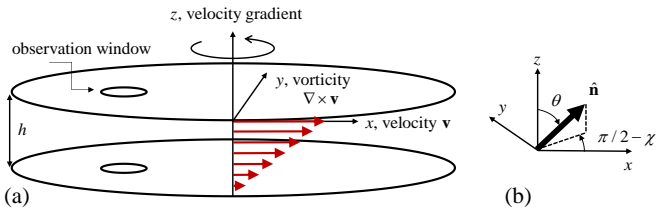

Fig. 2. (a) Principal scheme of the plate-plate rotating shearing cell in Linkam Optical Shearing System CSS450; in the actual device, it is the bottom disk that is rotating; (b) definition of the director tilts with respect to the flow direction and the gradient direction.

To reconstruct the director field and its dependencies on the applied shear, we used POM with optical compensators of two types. The first type is a static full-wave-plate (FWP) 530 nm optical compensator with the slow axis $\boldsymbol{\lambda}_g$ oriented at 45° to the crossed polarizers. A slab of a birefringent material viewed between the two crossed polarizers and an FWP shows interference colors that depend on the orientation of $\hat{\mathbf{n}}$. For a material of negative birefringence, such as DSCG [64], the textural color appears magenta when $\hat{\mathbf{n}}$ is along the polarizers, blue when $\hat{\mathbf{n}}$ is perpendicular to $\boldsymbol{\lambda}_g$ and yellow when $\hat{\mathbf{n}}$ is parallel to $\boldsymbol{\lambda}_g$.

The second optical compensator, called Abrio LC PolScope (Cambridge Research Incorporation), is tunable. The LC PolScope relies on a monochromatic illumination of the wavelength $\lambda = 546$ nm and represents a liquid crystal slab that changes optical retardance under an applied electric field. Once the image is taken for a few different settings of the compensator, numerical analysis allows the unit to map the optical retardance $\delta$ of the sample in the range (0 – 273) nm and the in-plane orientation of the optical axis $\hat{\mathbf{n}}$ [65]. For a tangentially anchored quiescent LCLC sample, $\delta = |n_e - n_o| h$. For a sample under steady-state shear flow, $\hat{\mathbf{n}}$ can vary in space, thus PolScope determines only an effective optical retardance that depends on the director tilt. Another limitation is that the PolScope textures represent an average over a time interval of 2 s needed to acquire the images. The static FWP compensator allows one to obtain single images much faster, being limited only by the image acquisition rate of the video camera. The shear device is mounted at POM Olympus BX40, equipped with a video camera AM Scope MU 2000 (10 frames per second) or at Nikon E600 microscope equipped with the LC PolScope[66].

The size of director domains formed at $\dot{\gamma} \geq 1\,s^{-1}$ is determined with an open-source software package Fiji2/ImageJ[67]. For weak shear rates, $\dot{\gamma} \leq 100\,s^{-1}$, the domains are selected from a stack of images using the Color Threshold function and analyzed using the Analyze Particles function. For high shear rates, $\dot{\gamma} \geq 100\,s^{-1}$, at which stripe textures develop, fast Fourier transform (FFT) is used to find the periodicity of the stripes. Details of the optical analysis of the domain textures are described in the Electronic Supplementary Information (ESI), Figs. S1-S4. All the calculations and plots are performed with custom-written codes in MATLAB (MathWorks) and Mathematica (Wolfram).

X-ray diffraction is widely used to characterize LCLCs in equilibrium [68-73]. It is expected that rheological small-angle x-ray scattering (rheo-SAXS) and rheological wide-angle x-ray scattering (rheo-WAXS) could also provide an excellent insight into the structure under the shear flows. The rheo-SAXS/WAXS experiments were performed at the 11-BM CMS beamline at National Synchrotron Light Source II using the same model Linkam Optical Shearing System CSS450. The beamline was configured for a collimated X-ray beam of size 0.2 mm by 0.2 mm and a divergence of 0.1 mrad by 0.1 mrad with energy of 17 keV ($\lambda$=0.073 nm). Mica windows were used on the Linkam stage to improve the transmission and reduce background scattering. The SAXS and WAXS signals were collected by Pilatus 2M and 300K with a distance of 2.8 m and 1.6 m away from the samples, respectively. The sample-to-detector distance was calibrated using a silver behenate calibration standard. The background scattering was collected from the optical rheometer with an empty gap. The exposure time for both SAXS and WAXS was 20 sec. The WAXS results, Fig. S5 in ESI, confirm that the shear does not affect the stacking distance 0.34 nm of the DSCG molecules in aggregates.

The SAXS pattern shows crescent-like peaks that identify the nematic phase. The SAXS patterns are analyzed in terms of the angular distribution of the scattering intensity. The intensity at a constant azimuthal angle $\chi$ is integrated over the momentum transfer range $q_m \pm \Delta q$ with $q_m$ being the momentum transfer value corresponding to the maximum intensity and $\Delta q$ is taken as the half-width at half maximum of the intensity vs $q$ peaks. In the experiments, $q_m \simeq 0.13\,\text{Å}^{-1}$ corresponds to the lateral distance $\sim 4.8$ nm between the DSCG aggregates that is noticeably larger than the diameter $\sim 3.2$ nm of the aggregates, in agreement with the previous studies [69, 71, 72, 74]; the reason is that the DSCG aggregates are separated by water, as schematically illustrated in Fig.1b.

### III. Results

#### A. Shear viscosity vs. shear rate

Figure 3 shows the effective shear viscosity $\eta$ measured in the cone-plate geometry as a function of the shear rate $\dot{\gamma}$, where $\eta \propto \dot{\gamma}^{n-1}$ and $n=1$ is corresponding to the Newtonian fluid behavior. The material is shear-thinning: $\eta$ decreases as $\dot{\gamma}$ increases, $\eta \propto \dot{\gamma}^{n-1}$, where $n<1$ indicates a negative slope of $\eta(\dot{\gamma})$

. The behavior is similar to the classic Onogi and Asada "three-region" scheme [75] that is a universal rheological signature of many liquid crystalline materials ranging from LCP [2, 76] to cellulose nanocrystals dispersions [77, 78]. In region I, at the smallest shear rates, approximately in the range $0.1\,\text{s}^{-1} < \dot\gamma < 1\,\text{s}^{-1}$, the decline of $\eta$ is the sharpest, $\eta \propto \dot\gamma^{-0.2}$, with $n$ reaching its smallest value of 0.8. Shear-thinning is reduced when the shear rate approaches $\dot\gamma = 1\,\text{s}^{-1}$. In Region II, $1\,\text{s}^{-1} < \dot\gamma < 10\,\text{s}^{-1}$, $\eta$ does not change much with $\dot\gamma$, $\eta \propto \dot\gamma^{-0.02}$ and $n \approx 1$, this is why this region is often called a (pseudo)-Newtonian. In Region III, $10\,\text{s}^{-1} < \dot\gamma < 10^3\,\text{s}^{-1}$, the material shows a moderate shear thinning, $\eta \propto \dot\gamma^{-0.1}$ with $n \approx 0.9$. A comparative optical and SAXS/WAXS analysis of the three regimes presented below demonstrates that the dynamic director patterns in them are very different from each other, being of a smooth log-rolling type in Region I, disclination-dominated in Region II and stripes-dominated in Region III.

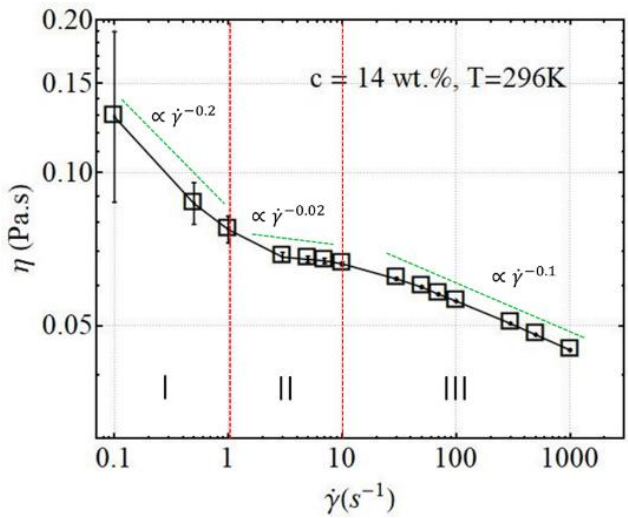

Fig. 3. Effective shear viscosity of the nematic DSCG, 14wt% dispersion in water, at 296 K. Region I: $0.1\,\text{s}^{-1} < \dot\gamma < 1\,\text{s}^{-1}$, Region II: $1\,\text{s}^{-1} < \dot\gamma < 10\,\text{s}^{-1}$; Region III: $10\,\text{s}^{-1} < \dot\gamma < 10^3\,\text{s}^{-1}$.

### B. Region I, $\dot\gamma < 1\,\text{s}^{-1}$, director alignment along the vorticity axis

The untreated substrates of the LCLC cells in both POM and SAXS-WAXS studies set a degenerate tangential alignment: in absence of the shear, the director $\hat{\mathbf{n}}$ can assume any orientation in the $(x, y)$ plane. To demonstrate this, we subjected the samples to shear at 10 s$^{-1}$ then stopped the shear and allowed the material to relax for ~15 min. The relaxation resulted in different initial orientations of the director in the plane of the sample, as shown in the left-hand columns of Fig.4 and Fig. 5. The purpose of different initial conditions is to uncover how universal are the features of director dynamics under weak ($\dot\gamma < 1\,\text{s}^{-1}$, Fig.4), and moderate ($1\,\text{s}^{-1} < \dot\gamma < 10\,\text{s}^{-1}$, Fig.5), shear rates.

The main effect of a weak shear, $\dot\gamma < 1\,\text{s}^{-1}$, applied at any angle to a local $\hat{\mathbf{n}}$, is to reorient the average director towards the vorticity $y$-axis, Fig. 4. This behavior, called a log-rolling regime, is characteristic of tumbling nematics [79, 80], and is also observed in LMWNs [4, 81], lyotropic [82, 83], and thermotropic [84, 85] nematic polymers. When $\hat{\mathbf{n}}$ is initially along the flow, the realignment time is $t \sim 10\,\text{min}$, Fig. 4a, which corresponds to the shear strain $\gamma = \dot\gamma\,t = \omega r t / h \approx 360$, where $\omega = 0.001\,\text{rad/s}$ is the angular velocity of the moving plate, $r = 7.5\,\text{mm}$ is the distance from the center of the shear stage to the observation window, and $h$ is the gap of the rheometer, thus $\omega r t \approx 4.5\,\text{mm}$ is the arc length distance traveled by the rotating disk with respect to the stationary disk at the point of observation. Shear-induced alignment along the vorticity axis is not perfect, as one often observes domains of typical scale (100-500) μm advected by the flow in which $\hat{\mathbf{n}}$ can be aligned in any direction, Fig. 4a,b, including the direction of flow. These domain textures are smooth and free of singular defects such as disclinations.

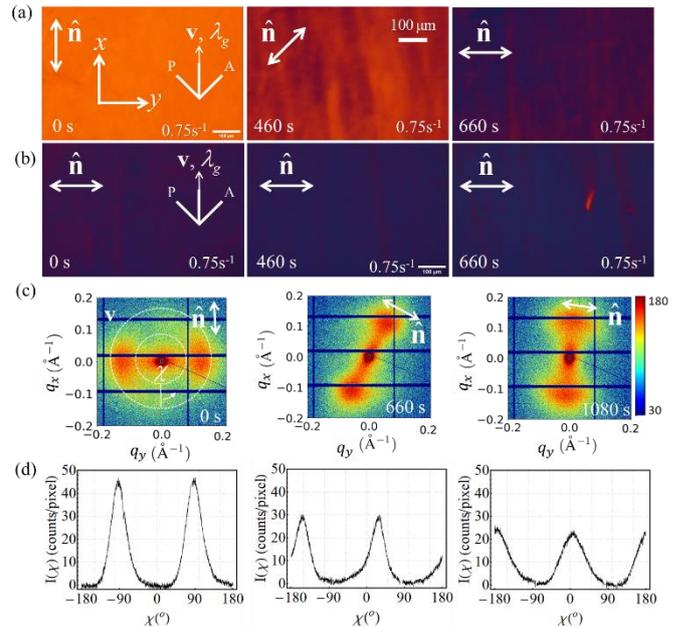

Fig. 4. Nematic structures in a 10 μm thick cell at $T = 296$K under a weak shear, $\dot\gamma = 0.75\,\text{s}^{-1}$, as revealed by (a,b) POM with crossed-polarizers and FWP and (c,d) SAXS. (a) Shear realigns the director from the initial direction along the flow towards the vorticity $y$-axis; time since the start of shear is shown at each texture; (b) director initially along the vorticity $y$-axis maintains its orientation. (c) SAXS patterns showing the nematic director projected onto-the $(xy)$ plane, and (d) the radially integrated SAXS intensity as a function of the azimuthal angle $\chi$, calculated as explained the methods section.

### C. Region II, $1\,\text{s}^{-1} < \dot\gamma < 10\,\text{s}^{-1}$, nucleation and dynamics of disclinations

The director field in this shear range is highly distorted, forming a polydomain texture regardless of the initial alignment, as evidenced by the textures documented in various modes of optical microscopy and by SAXS, Fig. 5,6. As described above, the samples were pre-sheared and allowed to relax. In Fig.5a, the initial alignment is close to the direction of flow, while in Fig.5b, the initial orientation is at some angle with respect to the flow. Once the shear starts, the director realigns, forming polydomain textures, shown in the central and right columns of Fig.5 for shear rates 1 s$^{-1}$ and 10 s$^{-1}$, respectively. In these experiments, as in other observations of shear-induced textures described below, the shear was applied for at least $\gamma = 100$ strain units before any textures were documented, in order to assure that the sample is in a stationary state, in which the average domain characteristics do not change with time. When viewed under POM in different settings, Fig. 5a,b,c, the polydomain texture represents small regions with $\hat{n}$ aligned along directions substantially different from the vorticity $y$-axis, including the direction of flow, as illustrated in Fig. 5c for $\dot\gamma \approx 1\,s^{-1}$. The domains, steadily advected by the flow, are elongated along the $x$-axis, so that their characteristic length $a_x$ is larger than their width $a_y$ along the vorticity axis, Fig. 5a. The length $a_x$ and width $a_y$ of the domains with $\hat{n}$ tilted away from the vorticity direction are determined by POM observations with FWP. The slow axis $\lambda_g$ makes an angle 45° with the vorticity axis so that the counterclockwise director tilts appear yellow and the clockwise tilts appear blue, as explained in the Supplement. The domains of a tilted $\hat{n}$ coexist with the remnants of the texture in which $\hat{n}$ is parallel to the vorticity axis, Fig. 6a,b. The director lines connecting one domain with the other follow a chevron trajectory rather than a sinusoid, Fig. 6a,b. The "zigs" and "zags" of the director tilts, of width $a_y$ each, are separated by walls of width $a_w$. The latter is measured in a POM setting with the slow axis of FWP oriented along the flow, so that the regions with the director parallel to the vorticity appear blue, see Supplement. All three parameters, $a_x$, $a_y$, and $a_w$, decrease with the increase of the shear rate in the range $1\,s^{-1} < \dot\gamma < 10\,s^{-1}$, Figs. 5a-d, approximately as $a_x \propto \dot\gamma^{-1\pm 0.05}$, Fig. 6c, and $a_y \propto a_w \propto \dot\gamma^{-0.5\pm 0.05}$, Fig. 6d. In the dependence $a_y = b\dot\gamma^{-0.5\pm 0.05}$, the proportionality coefficient is $b \approx 3.6\times 10^{-5}\,m\,s^{1/2}$.

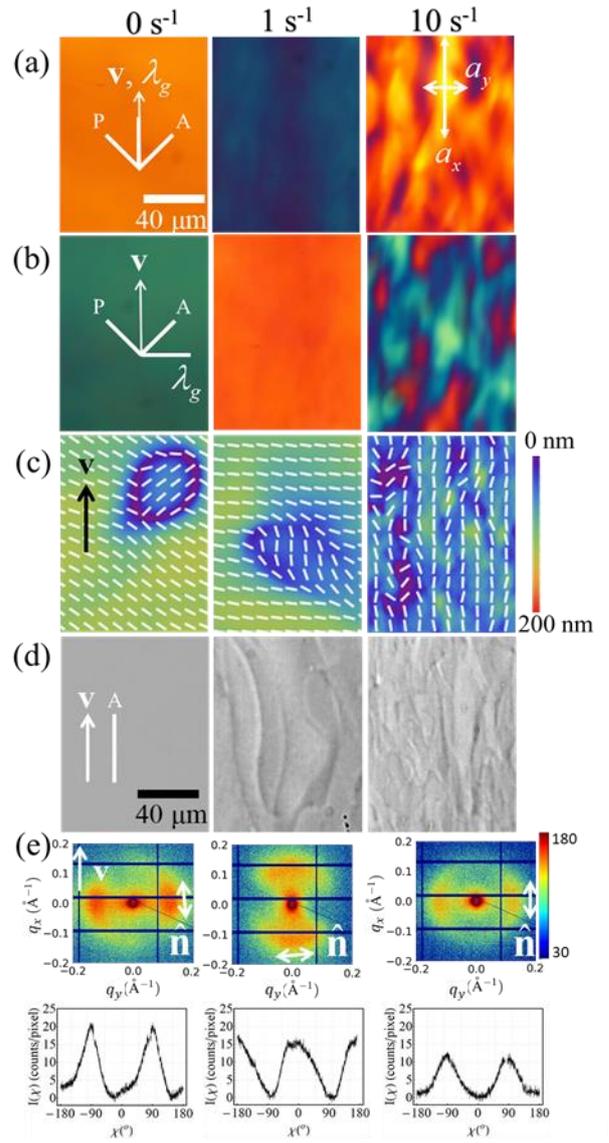

Fig. 5. (a-d) Optical microscopy and (e) SAXS textures of sheared nematic DSCG in the Region II of shear. (a,b) POM with the FWP's slow axis aligned (a) parallel to the flow direction, $\lambda_g \| v$, and (b) perpendicular to it, $\lambda_g \perp v$; (c) PolScope textures showing the local in-plane director pattern and the local optical retardance; (d) web of disclination loops observed with a single polarizer; (e) SAXS intensity distribution in the $x-y$ plane and the corresponding the radially integrated SAXS intensity as a function of the azimuthal angle $\chi$, calculated as explained the methods section. At $\dot\gamma = 1\,s^{-1}$, the average director shows a strong component along the vorticity axis, while at 10 s$^{-1}$ it is attracted to the shear plane. Sample thickness $10\,\mu m$ in parts (a-d) and $200\,\mu m$ in part (e).

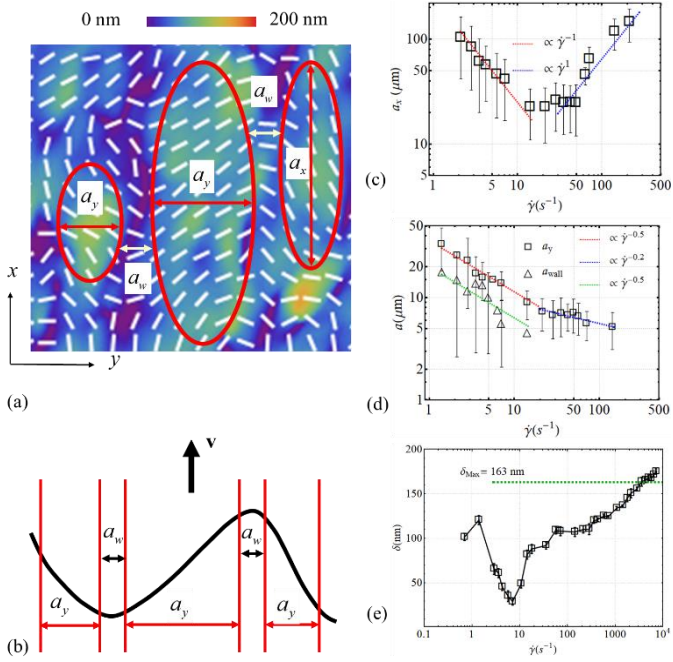

Figure 6. Textural parameters in Region II. (a,b) Schematic structure of the tilted-director domains in the chevron texture, of a length $a_x$ and width $a_y$, separated by walls of width $a_w$, in which the director is predominantly along the vorticity axis. Shear rate dependence of the characteristic (c) length $a_x$ and (d) width $a_y$ of the tilted director domains and walls $a_w$; (e) shear rate dependence of the effective retardance measured by PolScope; the retardance data are averaged over 2 s. All data are for a 10 μm thick sample.

Within Region II, the smooth polydomain texture is progressively replaced by a web of singular disclination loops of half-integer strength, Figs. 5,7,8. Nucleation of disclinations is a signature of tumbling nematics [7, 9, 86, 87]. A few isolated loops nucleate already at 1 s$^{-1}$, Fig. 7. The disclinations are topologically trivial as they appear from "nothing" and could collapse into a locally uniform structure; the loops are trivial when considered as the elements of the second homotopy group that describes point defects [5]. Fig. 7a-c show the time evolution of an isolated loop that first appears within the shear plane $(xz)$, and then realigns towards the $(xy)$ plane, being always elongated along the flow direction. Since the director outside the loop is along the vorticity $y$-axis and within the loop, it is along the $x$-direction of flow, the rotation vector $\mathbf{\Omega}$ of $\hat{\mathbf{n}}$ is parallel to the gradient $z$-axis, Fig. 7d. The normal $\mathbf{N}$ to the nucleating loop is initially along the vorticity axis, $\mathbf{N}=(0,1,0)$, but with time, it realigns towards the gradient direction, $\mathbf{N}=(0,0,1)$, Fig. 7.

The loop in the shear plane with $\mathbf{N}=(0,1,0)$ and $\mathbf{\Omega}||\hat{\mathbf{z}}$ is of a mixed twist-wedge type: the segments elongated along the flow are of a twist type, while the short segments perpendicular to the flow are of a wedge type [5, 9, 88-90], Fig. 7a,d. Nucleation of disclinations within the shear plane has been reported for sheared nematic polymers by De'Nève et al [9]. In our case, the loops often realign from the shear plane towards the $(xy)$ plane, as their open oval shapes clearly indicate that $\mathbf{N}$ is no longer along the vorticity direction, Fig. 7b,c. When $\mathbf{N}=(0,0,1)$, the loop can be of a mixed wedge-twist type ($\mathbf{\Omega}$ is not along the $z$-axis) and of a pure twist type, with $\mathbf{\Omega}||\hat{\mathbf{z}}$, Fig. 7e. Reorientation away from the shear plane and formation of twist loops can be explained by the smallness of the twist elastic constant in nematic LCLCs as compared to bend and splay constants [31]. However, the Region II textures show not only twist disclinations but also plenty of wedge segments perpendicular to the flow direction, Fig. 8a. These wedge segments are visualized by PolScope as "points" in the $(xy)$ plane of observation, around which $\hat{\mathbf{n}}$ rotates by $\pi$. If the rotation sense of $\hat{\mathbf{n}}$ coincides with that of circumnavigation around the core, the segment is of the +1/2 strength, if not, it is of the -1/2 type. In 3D, all these wedge segments and twist disclinations are topologically equivalent. The density of disclinations increases with $\dot{\gamma}$, Fig. 8b; their separation along the vorticity and velocity directions decreases, which correlates qualitatively with the data on shear rate dependence of $a_x$, $a_y$, and $a_w$ in Fig. 6c,d.

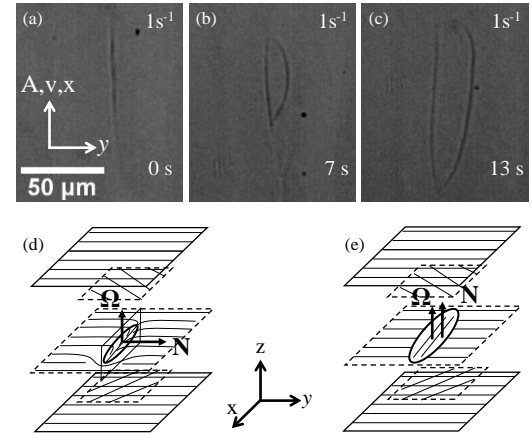

Fig. 7. Time sequence showing dynamics of a single disclination loop in a 10 μm thick sample during a steady shear at 1 s$^{-1}$; (a) isolated loop nucleates in the shear $(xz)$ plane and then (b,c) realigns towards the $(xy)$ plane; director distortions for (d) the mixed twist-wedge loop located in the shear plane, $\mathbf{N}=(0,1,0)$, and (e) the twisted loop located in the $(xy)$ plane, $\mathbf{N}=(0,0,1)$.

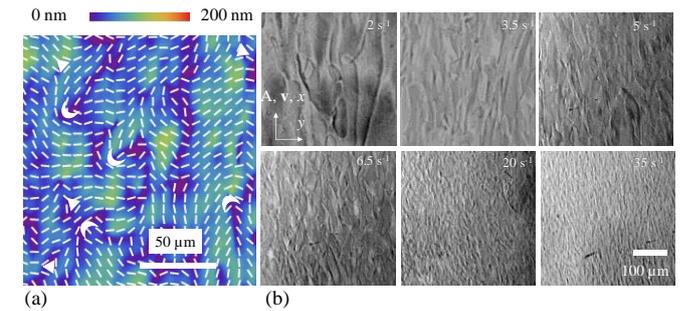

Fig. 8. Disclinations produced by shear in Region II: (a) PolScope texture showing wedge segments of disclinations in the director field projected onto the $(xy)$ plane; the four cores of -1/2 type are marked by triangles, the cores of four +1/2 disclinations are marked by crescents; shear rate 3 s$^{-1}$; (b) disclination density increases with the shear rate. 10 μm thick sample.

As the shear rate increases from 1 s$^{-1}$ to 10 s$^{-1}$, $\hat{\mathbf{n}}$ progressively turns towards the shear plane, as evidenced by the prevalence of the yellow interference color in Fig. 5a for $\dot{\gamma} = 10\,\text{s}^{-1}$, blue interference color in Fig. 5b, by optical axis orientation in Fig. 5c, and by SAXS signal, Fig. 5e. Figure 6e shows that the effective retardance $\delta$ in Region II reaches a minimum of about 30 nm, which is 5 times smaller than the actual retardance of a planar cell with the same thickness, the green dotted line in Fig. 6e, at the same temperature 296 K. One reason for a decreased $\delta$ is the nucleation of disclinations since their cores show a decreased scalar order parameter $S$ [91]. The decrease of $\delta$ can be also caused by partial realignment of $\hat{\mathbf{n}}$ along the velocity gradient $z$-axis, which would happen when the director tumbles in the shear plane. The third reason is possible twists of $\hat{\mathbf{n}}$ that are not accounted for in the working scheme of the PolScope, in which the optic axis is assumed not to twist around the propagation direction of light [65]. Thus the optical features point towards a complex 3D director pattern with a potential decrease of $S$ and with substantial projections of the director onto all three spatial axes, which is a signature of the tumbling behavior.

The middle of the Region II represents an important turning point in the scenario of director dependence on $\dot{\gamma}$. At $\dot{\gamma} < 5\,\text{s}^{-1}$, $\hat{\mathbf{n}}$ is predominantly along the vorticity axis, Fig. 9. Above $\dot{\gamma} \approx 5\,\text{s}^{-1}$, $\hat{\mathbf{n}}$ is predominantly in the shear plane, as manifested by the prevalence of the yellow colors in Fig. 9a and blue colors in Fig. 9b. Predominant alignment of $\hat{\mathbf{n}}$ along the flow at very high shear rates, in Region III, is clearly evidenced by optical and SAXS data in Figs.10-12 and by an increase of optical retardance in Fig. 6e.

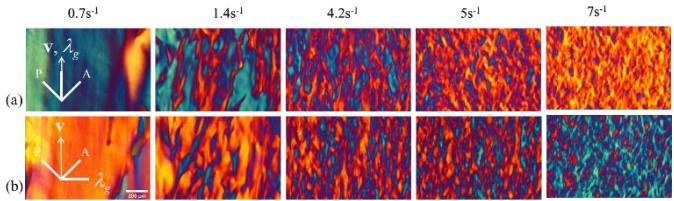

Fig. 9. Progressive realignment of the director projection from the vorticity axis at low shear rates $\dot{\gamma} < 5\,\text{s}^{-1}$ to the flow direction at higher shear rates, $\dot{\gamma} > 5\,\text{s}^{-1}$. The optic axis of the FWP is parallel to the flow in (a) and to the vorticity axis in (b). Yellow color in (a) indicates alignment along the flow, in (b) along the vorticity axis.

### D. Region II to Region III transition, $10\,\text{s}^{-1} < \dot{\gamma} < 500\,\text{s}^{-1}$, the transition from disclinations to stripes parallel to the flow

The most salient feature of the range $10\,\text{s}^{-1} < \dot{\gamma} < 500\,\text{s}^{-1}$ is a gradual elongation and transformation of disclination loops into stripes aligned along the flow, Fig. 10. The transition region also demonstrates a very strong increase of the optical phase retardance, Fig. 6e, indicating that $\hat{\mathbf{n}}$ aligns progressively better along the flow. SAXS data in Fig. 11 confirm that $\hat{\mathbf{n}}$ progressively realigns towards the flow at $\dot{\gamma} > 10\,\text{s}^{-1}$.

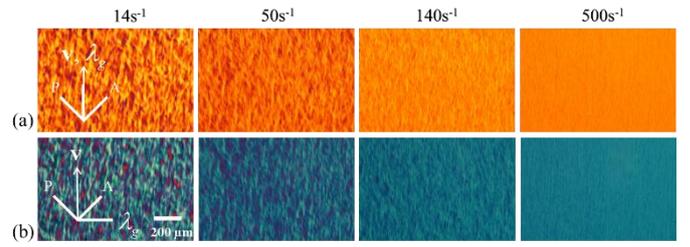

Fig. 10. The gradual replacement of the disclinations-infused polydomain textures into a texture of stripes observed in a POM with FWP with the increase of the shear rate (indicated in the figures) in 10 μm thick sample in the low-shear rate end of Region III. Note refinement of the domains at low shear rates and their elongation at higher shear rates. The director is predominantly realigned toward the shear flow (yellow color in part (a) and blue in part (b) at $\dot{\gamma} > 100\,\text{s}^{-1}$).

As the shear rate increases above ~20 s$^{-1}$, the domain width $a_y$ along the vorticity direction decreases $a_y \propto \dot{\gamma}^{-0.2 \pm 0.01}$, but less steeply than in the Region II. The long axes of the domains show a non-monotonous dependence on $\dot{\gamma}$. In Region II, $a_x$ decreases, $a_x \propto 1/\dot{\gamma}$, reaching a minimum $a_x \approx 25\,\mu\text{m}$ for $\dot{\gamma} = (10-40)\,\text{s}^{-1}$; at higher shear rates, $a_x \propto \dot{\gamma}$ increases dramatically, until it exceeds the size of the observation window; this elongation corresponds to a transformation from the texture of elongated disclinations to a texture of stripe domains.

### E. Region III, $\dot{\gamma} > 500\,\text{s}^{-1}$, stripes parallel to the flow.

SAXS, Fig. 11, and POM, Fig. 12, demonstrate the tendency of $\hat{\mathbf{n}}$ in Region III to align along the flow and replacement of the disclination web with stripe domains. In the stripe texture, $\hat{\mathbf{n}}$ is predominantly along the flow, with left and right tilts towards the vorticity axis, as shown schematically in Fig.12d. The SAXS data in Fig.11 show a small tilt (<5°) of the flow direction and the overall director from the "vertical" axis, which is an artefact of the experimental set-up, associated with a small shift (<0.5 mm) of the probing beam from the center of the observation window.

The small left and right periodic tilts of the director in stripe textures, Fig.12 a-f produce a weak POM contrast. To enhance it, the textures in Fig.12a-e are observed with the titled optical axis of the FWP. When the FWP optical axis is along the vorticity axis, Fig.12f, the overall blueish color in conjunction with the high optical retardance in Fig. 6e demonstrates clearly that the predominant director orientation is along the flow. The right and left director tilts are estimated by observing different orientations of the sample with the FWP to be about 5° for the shear rate 7100 s$^{-1}$. The period of the stripe-domains decreases with the shear rate, $p_{stripe} \propto \dot{\gamma}^{-0.1 \pm 0.07}$, Fig. 12h.

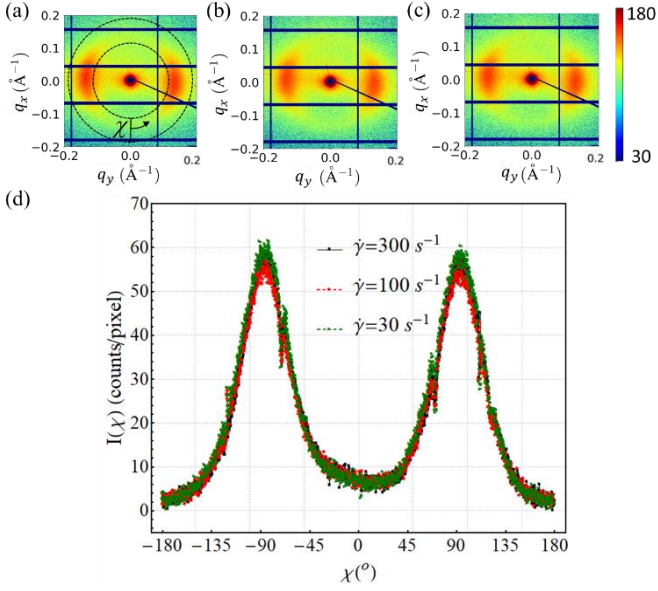

Fig. 11. SAXS patterns for a 200 μm thick sample demonstrating alignment of the director along the flow direction for shear rates (a) $30\,s^{-1}$ to (b) $100\,s^{-1}$ to (c) $300\,s^{-1}$, and (d) the radially integrated SAXS intensity in (a-c) as a function of the azimuthal angle $\chi$.

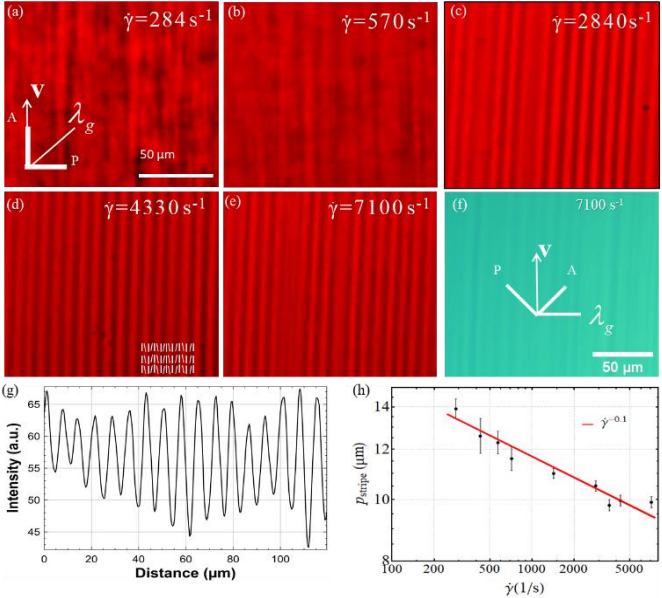

Fig. 12 (a-f) Stripe textures observed in 10.5 μm thick sample under strong shear with rates $\dot{\gamma} = (284 - 7100)\,s^{-1}$, observed with a tilted optical axis of the FWP compensator; (d) includes a schematic director map of stripes, (f) POM texture with the optical axis of the FWP compensator along the vorticity axis; $\dot{\gamma} = 7100\,s^{-1}$, (g) intensity of transmitted light in the POM texture in part (d); (h) period of stripes vs. shear rate.

In Region III, the effective retardance $\delta$ monotonously increases with $\dot{\gamma}$, until it reaches the value $\delta_{max} = 163$ nm that represents a retardance of a well-aligned quiescent monocrystal sample, $\hat{\mathbf{n}} = (1,0,0)$, at $\dot{\gamma} \approx 3200\,s^{-1}$. Shear rates above $3200\,s^{-1}$ produce retardance $\delta = 175$ nm that exceeds $\delta_{max}$ by 12 nm, Fig. 6e. A plausible explanation is that the high-rate shear enhances the scalar order parameter, presumably by favoring elongation of the domains that find the ends of each other easier.

## IV. Discussion

### A. Applicability of the Leslie-Ericksen model.

If the flow changes only the director field $\hat{\mathbf{n}} = \hat{\mathbf{n}}(x,y,z)$, but not the scalar order parameter, the rheological response can be described by the phenomenological Leslie-Ericksen theory with five independent viscosities, $i$ =1-5, known as Leslie coefficients and 3 bulk elastic moduli known as the Frank constants, $K_1$ for splay, $K_2$ for twist, and $K_3$ for bend. It is often assumed that the Leslie-Ericksen theory is applicable when the characteristic shear rate $\dot{\gamma}$ is smaller than the rotation diffusion coefficient $D_r = \kappa D_{r0} (\nu L^3)^{-2}$, where $\kappa$ is an empirical dimensionless coefficient, $D_{r0}$ is the rotary diffusivity, $\nu$ is the number density, and $L$ is the aggregates length[2]. $D_r$ can be roughly estimated by considering the aggregates as rods in a dilute dispersion with $D_{r0} = \frac{3k_B T}{\pi \eta_s L^3}\left(\ln\frac{L}{d} - 0.8\right)$, where $k_B$ is Boltzmann constant, $T$ is the temperature, $\eta_s = 9 \times 10^{-4}$ kg m$^{-1}$ s$^{-1}$ is the viscosity of solvent (water), $d$ is the aggregate diameter, and $\nu = 4\phi/\pi d^2 L$ is the number of aggregates per unit volume with $\phi$ being the volume fraction [2]. Electron microscopy of the 15 wt.% solution of DSCG shows $L$ is in the range 25-80 nm [92], while $d = 1.6$ nm [74]; for semi-flexible molecules, $\kappa \sim 10^4$ [2, 93]. The upper limit $L = 100$ nm yields $D_r \approx 3 \times 10^5\,s^{-1}$, which implies that for $\dot{\gamma} \leq 10^5\,s^{-1}$, the Leslie-Ericksen arguments might be applicable. However, this consideration overestimates the upper limit of $\dot{\gamma}$ at which the model is applicable, since it does not account for a strong coupling of director gradients and the scalar order parameter $S$ in LCLCs, as discussed below.

As demonstrated by Zhou et al [91], a peculiar property of the nematic LCLCs is that the scalar order parameter $S$ decreases significantly at the cores of singular disclinations that extend over unusually long scales, ~10 μm. A dramatic decrease of the optical retardance $\delta$ in the range $1\,s^{-1} < \dot{\gamma} \leq 100\,s^{-1}$, Fig. 6e, where the disclinations are abundant, might be associated with the decrease of $S$ at the disclination cores. Since in the range $1\,s^{-1} < \dot{\gamma} \leq 100\,s^{-1}$ the shear-induced textures feature multiple defects and domains, it is impossible to clearly separate the effect of director distortions over the sales longer than ~10 μm and the effect of a reduced $S$ at scales shorter than ~10 μm. On the other hand, a significant increase in optical retardance at the highest shear rates, Fig. 6e, together with a predominant alignment of the director along the flow in Region III, Figs.10-12, indicates a possibility of an enhanced $S$ triggered by the flow-induced merger of LCLC aggregates. Thus, in LCLCs, both the director and the scalar order parameter respond to the relatively low shear rates $\dot{\gamma} \geq 1\,s^{-1}$, once the disclinations start to proliferate.

## B. Region I, realignment along the vorticity axis.

If $\hat{n}$ is confined to the shear plane, then the viscous torque around the $y$-axis is defined by two Leslie coefficients [80] $\alpha_2$ and $\alpha_3$, $\Gamma_{visc} = (\alpha_3 \sin^2\theta - \alpha_2 \cos^2\theta)\dot{\gamma}$, where $\theta$ is the angle between the $z$-axis and $\hat{n}$, Fig. 2b. While $\alpha_2$ is always negative in calamitic nematics, $\alpha_3$ can be either negative or positive. When $\alpha_2/\alpha_3 < 0$ the viscous torque acting on $\hat{n}$ in the shear plane, is nonzero for any $\theta$. In a tumbling nematic, a small shear displacement applied to a planar sample with $\hat{n}$ parallel to the flow, will realign $\hat{n}$ in the same sense as the shear flow vorticity [94-96]; in the geometry of Fig. 1a, this is a counterclockwise rotation of $\hat{n}$. In absence of any other torques, the viscous torque will rotate $\hat{n}$ indefinitely [2]. Spatial confinement that sets $\hat{n}$ parallel to the bounding plates thanks to the so-called surface anchoring and elasticity of the nematic resist this scenario.

If $\hat{n}$ remains in the shear $(xz)$ plane, the elastic deformations are predominantly of splay-bend type. The relative importance of the viscous and elastic torques is expressed by the Ericksen number $Er = \dfrac{\eta \dot{\gamma} h^2}{K}$. Using the typical values of the elastic moduli of splay and bend [31], $K_1 \approx K_3 \approx 10\,\text{pN}$, cell thickness $h = 10^{-5}\,\text{m}$, and the effective shear viscosity [31] $\eta = 0.1\,\text{Pa s}$, one estimates $Er = \dfrac{\eta \dot{\gamma} h^2}{K} \approx [1\,\text{s}]\dot{\gamma}$. If the director is confined to the shear plane, the elastic forces will prevent tumbling until the Ericksen number reaches a critical value [80] $Er_{in} = 4.8\sqrt{|\alpha_2|/\alpha_3}$. On the other hand, the threshold of the out-of-plane director distortions is [80] $Er_{out} = 9.51\sqrt{K_{22}|\alpha_2|/K_{11}\alpha_3}$. In a typical LMWN, $K_{11}/K_{22} \sim 2-3$ and $Er_{out}/Er_{in} > 1$, thus the shear triggers tumbling in the shear plane *before* the director could realign along the vorticity axis. In 14wt% DSCG at 296 K, $K_{11}/K_{22} = 12$ [31] and thus $Er_{out}/Er_{in} = 0.6$. In other words, since $Er_{out} < Er_{in}$, the theory suggests that the nematic DSCG with a small twist constant should realign the director upon a weak shear towards the vorticity axis, as observed experimentally, Fig. 4. Further comparison with the experiment is difficult since the elastic and anchoring forces in cells with untreated plates are not well defined.

The experimental data in Fig. 4 and the theoretical estimates that suggest director deviations away from the shear plane are supported by the numerical simulations performed by Tu et al [97] for a nematic with a small twist elastic constant, $K_{11}/K_{22} = K_{33}/K_{22} = 10$, which is close to the experimental data for the explored DSCG [31]. In these simulations, the director is demonstrated to gradually realign from the initial orientation along the $x$-axis to the vorticity axis, within about 35 strain units. During this realignment, the director also rotates around the vorticity axis, i.e., it engages in a kayaking motion.

## C. Mechanism of tumbling

A qualitative molecular picture that explains why DSCG is a tumbling nematic, as opposed to most of thermotropic LMWN that are flow-aligning is presented in Fig. 13. The scheme compares collisions of molecules in LMWN, Fig. 13a, and of aggregates in LCLCs, Fig. 13b. In LMWNs, the molecules are of an ellipsoidal shape and the collision forces, shown as $\mathbf{f}_{coll}$ form a positive torque $\Gamma_{visc} > 0$ so that $\alpha_3 < 0$. In the LCLCs, the building units are of a cylindrical shape with the ends that are perpendicular to the axis of aggregates. Thus the collision forces with the neighboring molecules that move into opposite directions above and below a given aggregate, produce a clockwise torque $\Gamma_{visc} < 0$ that implies $\alpha_3 > 0$, Fig. 13. This idea is similar to the one put forward by Helfrich to explain why $\alpha_3 > 0$ in the proximity of the nematic-to smectic A phase transition [98]. Note that DSCG nematics can show structural stacking fault defects such as Y-junctions and C-defects [31]. Both these configurations will support the scheme in Fig. 13b and contribute to the positive value of $\alpha_3$.

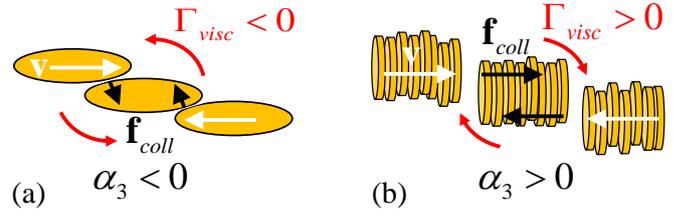

Fig. 13. Scheme explaining $\alpha_3 < 0$ in regular low-molecular nematics and $\alpha_3 > 0$ for LCLCs formed by cylindrical aggregates.

## D. Balance of viscous and elastic forces in Region II.

The details of the dynamic textures are determined by the balance of the viscous torque and the Frank-Oseen elastic torque associated with the gradients of $\hat{n}$. Assuming smooth director reorientation from one domain to the next along the vorticity axis over some characteristic scale $a_y$, one usually estimates the elastic energy density as $\sim K/a_y^2$, see Refs. [99-101]. The elastic energy should be comparable to the viscous work $\sim \eta \dot{\gamma}$ that creates the distortions; the balance results in the prediction $a_y \sim \sqrt{K/\eta\dot{\gamma}} \propto \dot{\gamma}^{-1/2}$ [99, 100]. The experimental dependence $a_y = b\dot{\gamma}^{-0.5\pm 0.05}$ in Fig. 6d confirms this scaling, but only in Region II. The coefficient $b$ should be of the order of $\sqrt{K/\eta}$. Choosing the average elastic constant [31] $K = 10\,\text{pN}$ and the shear viscosity $\eta = 0.1\,\text{Pa s}$, one estimates $b \sim \sqrt{K/\eta} \sim 10^{-5}\,\text{m s}^{1/2}$, while the experimental value $b = 3.6 \times 10^{-5}\,\text{m s}^{1/2}$. Given uncertainties and the absence of numerical coefficients, the agreement is reasonable. Moreover, it can be improved by noticing that the energy influx associated with the increasing shear rate can also be absorbed at a length scale smaller than $a_y$, such as the width of the domain walls, $a_w \approx a_y/2$ in Fig. 6d, separating the left and right tilts of the director in the chevron textures, Fig. 6a,b. The width $a_w$ can be associated with the walls parallel to the $z$-axis, as in Fig. 6a, but also with the walls perpendicular to the $z$-axis which are hard to observe. The elastic energy per unit volume of the cell with chevron-like domain stores predominantly at the ridges and scales as $K/a_w a_y$, rather than as $K/a_y^2$. This scaling would increase $b$ in the dependency $a_y = b\dot{\gamma}^{-0.5}$, yielding $b = \sqrt{Ka_y/\eta a_w} \approx 1.4\sqrt{K/\eta}$ for $a_y/a_w \approx 2$, Fig. 6d.

Comparing the elastic energy density $K/a_w a_y$ to the shear work density $\sim \eta\dot{\gamma}$, and using $K = 10$ pN, $\bar{\eta} = 0.1$ Pa s, $a_y = 10$ μm, one estimates that the width of the walls $a_w \sim K/(a_y \eta \dot{\gamma})$ decreases from $a_w = 10$ μm at $\dot{\gamma} = 1$ s$^{-1}$ to a submicron $a_w = 0.1$ μm at $\dot{\gamma} = 100$ s$^{-1}$, which is below the resolution limit of POM. Interestingly, the existence of two different length scales, $a_w$ and $a_y$, along the vorticity axis is also supported by numerical simulations of sheared LCPs [11, 102] in which the relatively wide "zigs" and "zags" of the director tilt and periodicity $a_y$ are separated by very narrow regions of a width $a_w < a_y$. Yang et al [102] called these regions "cigar-shaped domains" at tips of which the disclinations nucleate in numerical simulations. Note that in Region II, where $a_y \propto \dot{\gamma}^{-1/2}$, one would expect the same shear rate scaling of the domain walls' width $a_w \sim K/(a_y \eta \dot{\gamma}) \sim \dot{\gamma}^{-1/2}$, which agrees with the dependency in Fig. 6d. Below we use elasticity arguments to show that the decrease of $a_w \sim K/(a_y \eta \dot{\gamma})$ as the shear rate increases is the reason for the replacement of smooth polydomain texture with the disclination loops.

### E. Nucleation of disclinations in Region II.

The elastic energy density of the walls $f_{wall} \approx K/a_w a_y$ in the polydomain textures should be contrasted to that of the singular disclination loops which scale as $f_{discl} \approx \pi K \ln(a_y/r_c)/4a_y^2$, where $r_c$ is the core radius of the disclination and $a_y$ is the typical separation between the disclinations. As already discussed, the core radius of disclinations in LCLCs is large [91], thus in the order of magnitude, $\pi \ln(a_y/r_c)/4 \sim 1$. When $\dot{\gamma}$ increases and $a_w$ approaches $r_c$ from above, the elastic energy $f_{wall}$ stored at the chevron ridges raises. The ratio $f_{discl}/f_{wall} \sim \pi a_w \ln(a_y/r_c)/4a_y$ can be rewritten, using the relationship $a_w \sim K/(a_y \eta \dot{\gamma})$, as $f_{discl}/f_{wall} \sim \pi K \ln(a_y/r_c)/(4\eta \dot{\gamma} a_y^2)$. As the shear rate increases, at some point the ratio $f_{discl}/f_{wall}$ becomes smaller than 1 and the elastic stress at the walls is released by nucleating disclination loops. This condition is fulfilled at $\dot{\gamma} > \dot{\gamma}_{discl}$, where $\dot{\gamma}_{discl} \approx \frac{\pi \ln(a_y/r_c)}{4} \frac{K}{\eta a_y^2}$, estimated to be about $\sim 1$ s$^{-1}$ with the typical parameters presented above and $\pi \ln(a_y/r_c)/4 \sim 1$. The estimate $\dot{\gamma}_{discl} \sim 1$ s$^{-1}$ is in a very good agreement with the experimental data for the nucleation and proliferation of disclinations that starts at $\dot{\gamma} \geq 1$ s$^{-1}$, Figs. 5, 7-9.

### F. Periodic stripes in Region III.

In Region III, the disclinations are replaced by the stripes parallel to the flow. The predominant director orientation is also parallel to the flow, with periodic left and right tilts towards the vorticity direction. Apparently, the flow enhances alignment along the x-direction and facilitates elongation of aggregates by connecting neighboring ends, which might explain a strong increase of phase retardation in Fig. 6e. The stripes are of the chevron type, Fig. 12d, of width $a_y \leq 10$ μm, which decreases as $\dot{\gamma}$ increases. Besides the evidence of left and right tilts of the director that produce the optical contrast of the stripes, other details are hard to decipher because of the limited resolution of POM. We expect that the increase of $\dot{\gamma}$ would increase the director gradients in the regions separating left and right tilts, similarly to the case of $a_w$ discussed above.

## Conclusions

We explored the viscoelastic and structural response of the nematic LCLC DSCG to the simple shear flow. The effective shear viscosity of DSCG shows a three-region behavior as the shear rate increases, similar to the behavior of liquid crystal polymers. In Region I, $\dot{\gamma} \leq 1$ s$^{-1}$, the material shows a pronounced shear thinning, while the texture shows large domains tending to realign along the vorticity axis, which is most likely assisted by the small value of the twist elastic constant of DSCG. In Region II, $1$ s$^{-1} \leq \dot{\gamma} \leq 10$ s$^{-1}$, the viscosity is nearly constant, while the director tilts away from the vorticity axis. The optical texture can be qualitatively presented as domains of tilted director separated by regions in which the director is still along the vorticity axis. The domains are elongated along the flow. Their width $a_y$ and length $a_x$ shrink as the shear rate increases. The director trajectories transversing the domains resemble chevrons with straight regions of width $a_y$ and sharp ridges-walls of a width $a_w < a_y$ that decreases as $\dot{\gamma}$ increases. The elastic response of the system to the increased shear is through the decrease of $a_y$ and $a_w$, which results in the accumulation of high elastic energy at the chevron ridges. At some point, $\dot{\gamma} \geq \dot{\gamma}_{discl} \approx \frac{\pi \ln(a_y/r_c)}{4} \frac{K}{\eta a_y^2} \sim 1$ s$^{-1}$, the director gradients at the chevron ridges become too strong and are relieved by nucleation of disclination loops. The disclinations replace director orientation along the vorticity axis with the director orientation along the flow direction. The loops appear first in the shear plane but realign along the vorticity-flow plane, which we relate to a small value of the twist elastic constant in DSCG. The density of the disclination loops increases with the shear rate. In Region III with a moderate shear-thinning behavior, the web of disclinations elongates along the flow direction, as the length $a_x$ of the domains increases with $\dot{\gamma}$ until they become longer than the field of view. The disclinations are eventually replaced by a stripe texture in which the director is predominantly along the flow with alternating stripes of left and right tilts. The observed behavior points to the tumbling character of the DSCG nematic with the prevalent director orientation that can be either along the vorticity direction at low shear rates or along the flow at high shear rates.

## Conflicts of interest

There are no conflicts to declare.

## Acknowledgements

The work is supported by NSF grant DMREF-DMS-1729509. This work used the 11-BM CMS beamline of National Synchrotron Light Source II (NSLS-II), Brookhaven National Laboratory (BNL), a U.S. Department of Energy User Facility operated for the Office of Science by BNL under Contract DE-SC0012704. This research was completed while HB and ODL participated in KITP Active 20 program, supported in part by the NSF grant PHY-1748958 and NIH grant R25GM067110.